# *What are we talking about when we discuss the Born-Oppenheimer approximation?*

Olimpia Lombardi, Sebastian Fortin, Juan Camilo Martínez González, and Hernán Accorinti
*Universidad de Buenos Aires, CONICET*

## 1.- Introduction

The Born-Oppenheimer Approximation (BOA) is a widely used strategy in quantum chemistry due to its efficacy in its validity scope. Originally proposed by Max Born and J. Robert Oppenheimer (1927) to calculate molecular energy levels, the BOA is now formulated in a substantially different way, and is useful to explain other molecular properties.

Recently, Nick Huggett, James Ladyman, and Karim Thébault (2024) published an extensive article (hereinafter referred to as HLT) discussing the BOA. Their primary aim is to express a strong disagreement with our position about the matter, according to which the BOA includes a classical assumption that is incompatible with the Heisenberg Principle. The authors, by contrast, argue that the BOA requires no classical assumption, suggesting that the reduction of chemistry to physics is thereby ensured.

The aim of the present work is to critically examine the HLT article in order to explain why we find its arguments unconvincing. With this purpose, the article is organized as follows. Section 2 analyzes some of the argumentative strategies used in the HLT article that we consider weak, since involving a non-sequitur, or even problematic, as they misrepresent the arguments they criticize. Section 3 addresses technical issues central to the debate, focusing on the conceptual implications of the so-called Clamped Nuclei Approximation, on the content of Brian Sutcliffe and Guy Woolley's paper (2012a), supposedly responded to by Thierry Jecko (2014), and on the approximation strategy used by Jecko in his paper. Section 4 clarifies the conceptual framework underpinning our perspective on BOA, which confers our analysis its philosophical significance—this framework is only briefly mentioned and insufficiently examined in the HLT article. Finally, the Closing Remarks make explicit the philosophical view about science implicitly underlying the HLT article.

## 2.- Argumentative issues

Given the central role of argumentation in philosophy, a thorough analysis of the reasoning employed in any philosophical work is indispensable. In this section, we undertake this analysis.

*2.1.- About scientific practice*

The HLT article suggests that the position of those who find classical assumptions in the BOA lacks a solid foundation in scientific practice. According to the article, we assert that quantum chemistry and quantum physics are explicitly "in conflict" (HLT 2024: 1, 3, 7). For this reason, it is considered necessary to stress that "[t]he *conflict* with the Heisenberg uncertainly relation of Clamped *is irrelevant to the use* of models based on BO." (HLT 2024: 8, emphasis added). In this way, the HLT article installs the idea that we conceive our arguments to be *relevant to the use* of BO-based models in scientific practice. Unfortunately, this is an inaccurate reading of our work.

Our arguments concerning the BOA are framed in the context of the problem of reduction—not the general reduction of chemistry to physics, but rather *the strict reduction of molecular structure to quantum mechanics* (this point will be discussed in Section 4). Specifically, the question is whether molecular structure can be explained exclusively in non-relativistic quantum terms that could support the ontological reduction of molecular structure as understood in chemistry. Of course, one might ask what is meant by ontological reduction. One might also appeal to a theory "more fundamental" than quantum mechanics, or even to a future "final theory" that would explain all aspects of reality in physical terms. However, we are not addressing these broader issues here. The relevant point for us is the role of the BOA in the claims about the reduction of molecular structure: *we never questioned its practical applications*.[1] Indeed, we explicitly state that "[t]he *use* of the Born-Oppenheimer approximation is pervasive in quantum chemistry, and this fact *must not be questioned*: its justification relies on its own success." (Fortin and Lombardi 2021: 384; emphasis added).

*2.2.- Claims about the Heisenberg principle.*

The HLT article states: "To our knowledge, it has never been claimed in the scientific literature that BO violates Heisenberg uncertainty." (2024: 4). Let us take a closer look at this statement.

The paper by José Luis Villaveces and Edgar Daza (1990), referenced in Note 3 of the HLT article, is certainly a scientific work that specifically addresses the technical issues raised by the attempts to define or explain molecular structure in quantum terms—the context of our discussion. In particular, the authors say: "In most of quantum chemical literature a structure is identified with a single point in the nuclear coordinate space $^mR$. This model is in contradiction with Heisenberg uncertainty principle, since it gives a unique and well determined position to each nucleus; besides, it also impoverishes the chemical intuitive concept which has an enormous heuristic potentiality." (1990: 100-101). Other claims by scientists have been overlooked in the HLT article. For example: "Freezing fixes a well-determined value $\vec{R}_\alpha$ for the α-th position observable $\hat{Q}_\alpha$. On the other hand, $\hat{T}_K = 0$, so that the α-th momentum observable $\hat{P}_\alpha$ also has a sharp value, while in quantum mechanics the Heisenberg uncertainty relation $\Delta Q \cdot \Delta P \geq \hbar/2$ must be fulfilled." (Primas and Müller-Herold 1990: 154); "[t]he claim that molecules exist in an absolute sense is, in any case, incompatible with the fundamental principles of quantum mechanics." (ibid. 318). Or, more recently, in the discussion about the BOA it is explicitly said: "specifying precisely the positions $\{\mathbf{t}^n\}$ for stationary nuclei violates Heisenberg's uncertainty principle." (Woolley 2022: 281-282).

---

[1] In this sense, we endorse Roberto Torretti's view that "it is pragmatic realism, not the nostalgic kryptotheology of «scientific realism», that best expresses the real facts of human knowledge and the working scientist's understanding of reality." (Torretti 2000: 115).

In discussing the 2012a paper by Sutcliffe and Woolley, the HLT article adds: "at no point do they [Sutcliffe and Woolley] claim that BO violates the Heisenberg uncertainty principle." (2024: 9). Although they do not explicitly use the expression "BOA violates the Heisenberg uncertainty principle", it is difficult not to draw this conclusion from the statement quoted in the HLT article itself a few lines earlier: "An extra choice of fixed nuclear positions must be made to give any discrete spectrum and normalizable eigenfunctions. In our view this choice, that is, the introduction of the clamped-nuclei Hamiltonian, by hand, into the molecular theory is the essence of the «Born-Oppenheimer approximation»" (Sutcliffe and Woolley 2012a: 7). Although in this case Sutcliffe and Woolley address the issue arising from the continuous spectrum of the Electronic Hamiltonian (we will come back to this point in the next subsection), this claim is in resonance with their broader body of work, from Woolley's classic 1978 paper "Must a molecule have a shape?" to more recent publications such as "Is chemistry really founded in quantum mechanics?" (Sutcliffe and Woolley 2022). These examples are just a fraction of the authors' extensive contributions on this area. Moreover, Sutcliffe and Woolley are not alone in highlighting this issue. For instance, Anton Amann and Ulrich Müller-Herold ask: "How is it possible for quantum mechanics to be a theory of chemistry if the way in which chemistry states its results contradicts quantum mechanics?" (1999: 50). A recent scientific paper remarks that the BOA "in essence, imposes the molecular-structure concept upon quantum theory by explicitly treating the atomic nuclei as fixed classical point charges, eventually leading to the molecular-orbital theories that elegantly rationalize observed chemical phenomena. The BOA thus is highly successful, and importantly, its mathematical validity and numerical accuracy is very well established. The fundamental conflict persists, however." (Lang et al. 2024: 1760).

Summing up, the HLT article dismisses the first group of papers because they do not explicitly include the terms "Born-Oppenheimer approximation," and the second group of works because they do not use the words "violates Heisenberg principle." Only by adopting this strictly literal reading of the scientific literature can the HLT article characterize our view about the tension between the BOA and quantum mechanics as a thesis that has never been defended in science. This literal approach may also explain why the HTL paper does not extend its criticisms to Robin Hendry, who, although not a practicing chemist, is one of the leading contemporary philosophers of chemistry. It is true that Hendry does not explicitly state that the BOA violates the Heisenberg principle, but many of his claims about BOA-models emphasize that they "simply assume the facts about molecular structure that ought to be explained" (Hendry 2010: 186); in other words, "when tractable quantum-mechanical methods are applied to complex molecules, it seems that the semi-classical nature of the molecule is *presupposed*." (Hendry 1998: 135; emphasis in the original).

The HLT article continues: "So either philosophers of science have uncovered an important scientific fact that scientists themselves have somehow missed, or the arguments of the philosophers are incorrect." (2024: 4). First, the incompatibility between the BOA and quantum mechanics is not a scientific fact in the sense that it impacts scientific work: as remarked above, it does not hinder the use of BO-based models in the practice of chemistry. Second, as just argued, suggesting that scientists have missed such incompatibility arises from an overly literal or partial reading of the scientific literature (further scientific claims will be considered in Section 4).

*2.3.- From particular to general*

The HLT paper argues that, contrary to the prima facie assumption that quantum chemistry exemplifies the success of reductionism, some philosophers of chemistry hold a non-reductionist view of quantum chemistry: "A principal argument for this view is based upon a supposed *non-quantum* features of the 'Born-Oppenheimer approximation' (BO)" (2024: 2). The article seeks to challenge anti-reductionist arguments by arguing that there is no genuine incompatibility between the BOA and quantum mechanics.

This argumentative strategy draws a general conclusion about reduction based on a single case—the BOA. Let us suppose, for the sake of the argument, that the BOA does not involve classical assumptions. The general conclusion would follow if the BOA were the only manifestation of the tension between quantum chemistry and quantum mechanics. However, as we have emphasized in our previous papers, the BOA is not the primary argument for adopting a non-reductionist stance in this field. On the contrary, we explicitly state that the main obstacle to reduction is the so-called "symmetry problem": "if the interactions embodied in the Hamiltonian of the molecule are Coulombic, the solutions of the Schrödinger equation have certain symmetries that cannot account for the asymmetries of the molecular structure." (Fortin, Lombardi, and Martínez

González 2018: 133; see also interview in Scerri 2023). This problem is particularly serious if the existence of isomers is to be explained (Fortin and Lombardi 2021).

We are neither the only nor the first authors to point out the symmetry problem; in fact, it has been thoroughly examined by Sutcliffe and Woolley in their many works on the topic (see, e.g., Woolley and Sutcliffe 1977, Sutcliffe and Woolley 2012a, 2012b, 2022), clearly pointed out by the famous physicist and quantum chemist Per-Olov Löwdin (1989), and widely discussed in the field of the philosophy of chemistry by Hendry (e.g. 1998, 2010, 2022), just to mention some of the many examples. The crucial point to emphasize here is that *the symmetry problem is independent of the BOA* in the sense that it arises from the full, non-approximate Coulomb Hamiltonian of the molecule.

We will return to the symmetry problem in Section 4. For now, what has been discussed suffices to show why particular claims about the BOA are insufficient to draw general conclusions regarding the reduction of quantum chemistry to quantum physics.

## 3.- Technical issues

The discussion of the HLT article involves technical issues related to quantum chemistry and quantum mechanics that require detailed analysis. This section is devoted to that task.

*3.1.- The structure of the Born-Oppenheimer Approximation*

The BOA was originally proposed by Born and Oppenheimer in 1927 to calculate the energy levels of molecules, due to the fact that the time-independent Schrödinger equation (the eigenfunction-eigenvalue equation of the Hamiltonian) cannot be solved analytically. In that version, the full Hamiltonian $\hat{H}$ of a generic molecule is given by

$$\hat{H} = \hat{T}_e + \hat{U} + \hat{T}_N = \hat{H}_0 + \hat{T}_N, \tag{1}$$

where $\hat{T}_e$ is the kinetic energy of the electrons, $\hat{T}_N$ is the kinetic energy of the nuclei, and $\hat{U}$ is the potential energy of the system due to the Coulomb interactions. The Hamiltonian $\hat{H}_0 = \hat{T}_e + \hat{U}$ is usually called *Electronic Hamiltonian*.[2] The fundamental idea of Born and Oppenheimer is that the spectrum of $\hat{H}$ can be computed by conceiving $\hat{T}_N$ as a small perturbation of $\hat{H}_0$, with expansion parameter $\kappa$, defined in terms of the ratio between the electronic mass $m$ and the nuclear mass $M_0$ as $\kappa^4 = m/M_0$, such that $\hat{T}_N = \kappa^4 \hat{H}_1$:

$$\hat{H} = \hat{H}_0 + \kappa^4 \hat{H}_1. \tag{2}$$

---
[2] Let us stress that, although it is usually called 'Electronic Hamiltonian,' $\hat{H}_0$ depends on the position operators of the nuclei.

At this point, the authors say that, by setting $\kappa = 0$, the Schrödinger equation for $\hat{H} = \hat{H}_0$ becomes a differential equation that only depends on the electronic coordinates and where the nuclear coordinates appear as parameters: "Evidently, this represents the electronic motion for stationary nuclei." (Born and Oppenheimer 1927: 462). However, this is not evident at all: as will be shown below, $\hat{H}_0$ is not the Hamiltonian for fixed nuclei. The argument based on the disparity of masses, which may seem reasonable in a perturbative model that takes $(m/M_0)^{1/4}$ as its expansion parameter, is later transferred to other versions of the BOA (this point will be discussed in the next subsection).

The original version of the BOA is not the one typically presented in current literature. At present, the BOA is usually formulated in the following terms (Born 1951, Born and Huang 1954). Let us consider a molecule with (i) $N$ electrons, with charge $-e$ and mass $m$, whose positions and momenta are represented by the operators $\hat{r}_i$ and $\hat{p}_i$, respectively, and (ii) $A$ nuclei, with charge $+Z_g e$ and masses $M_g$, $g = 1, \ldots, A$, whose positions and momenta are represented by the operators $\hat{R}_g$ and $\hat{P}_g$, respectively. The quantum *Coulomb Hamiltonian* $\hat{H}$ of the molecule is given by

$$\hat{H} = \sum_{g=1}^{A} \frac{\hat{P}_g^2}{2M_g} + \sum_{i=1}^{N} \frac{\hat{p}_i^2}{2m} + \frac{e^2}{8\pi\varepsilon_0} \sum_{\substack{g,h=1 \\ g \neq h}}^{A} \frac{Z_g Z_h}{|\hat{R}_g - \hat{R}_h|} + \frac{e^2}{8\pi\varepsilon_0} \sum_{\substack{i,j=1 \\ i \neq j}}^{N} \frac{1}{|\hat{r}_i - \hat{r}_j|} - \frac{e^2}{4\pi\varepsilon_0} \sum_{g=1}^{A} \sum_{i=1}^{N} \frac{Z_g}{|\hat{r}_i - \hat{R}_g|}, \qquad (3)$$

where the first term is the nuclear kinetic energy $\hat{T}_N(\hat{P}_g)$, the second term is the electronic kinetic energy $\hat{T}_e(\hat{p}_i)$, the third term is the potential $\hat{V}_{NN}(\hat{R}_g)$ due to the interactions between the nuclei, the fourth term is the potential $\hat{V}_{ee}(\hat{r}_i)$ due to the interactions between the electrons, and the fifth term is the potential $\hat{V}_{eN}(\hat{r}_i, \hat{R}_g)$ due to the interactions between the electrons and the nuclei. The Electronic Hamiltonian $\hat{H}_0$ is $\hat{H} - \hat{T}_N$:

$$\hat{H}_0 = \sum_{i=1}^{N} \frac{\hat{p}_i^2}{2m} + \frac{e^2}{8\pi\varepsilon_0} \sum_{\substack{g,h=1 \\ g \neq h}}^{A} \frac{Z_g Z_h}{|\hat{R}_g - \hat{R}_h|} + \frac{e^2}{8\pi\varepsilon_0} \sum_{\substack{i,j=1 \\ i \neq j}}^{N} \frac{1}{|\hat{r}_i - \hat{r}_j|} - \frac{e^2}{4\pi\varepsilon_0} \sum_{g=1}^{A} \sum_{i=1}^{N} \frac{Z_g}{|\hat{r}_i - \hat{R}_g|}, \qquad (4)$$

The time-independent Schrödinger equation for $\hat{H}_0$ is[3]

$$\hat{H}_0(\hat{r}_i, \hat{R}_g) \Psi_n(r_i, R_g) = E_n \Psi_n(r_i, R_g), \qquad (5)$$

where the $\Psi_n(r_i, R_g)$ are the eigenfunctions of the Electronic Hamiltonian $\hat{H}_0$, that is, the stationary wavefunctions of the molecule ($\Psi_0$ is the ground state), and the $E_n$ are its energy levels.[4]

---

[3] For simplicity, hereinafter we will omit the dependence of the Hamiltonians on the electron momenta.

[4] At this point, the scientific literature introduces a change of variables to refer the Coulomb Hamiltonian to the center of mass of the system in order to abstract from some of its invariances. However, since this step does not have conceptual relevance for our discussion, here we will skip it for simplicity.

The first step of the BOA is the so-called *Clamped Nuclei Approximation*, which consists in conceiving the nuclei as fixed at definite positions, as implicitly assumed by Born and Oppenheimer in their original version. Mathematically, this approximation is performed on the Electronic Hamiltonian *by turning the operators* $\hat{R}_g$, which represent the quantum nuclear positions, *into classical vectors* $R_g$, playing the role of parameters that represent definite nuclear positions. On the basis of this assumption, the potential $\hat{V}_{NN}(\hat{R}_g)$, due to the interactions between the nuclei, becomes a constant scalar value that shifts the eigenvalues of the new Hamiltonian only by some constant amount and, therefore, can be neglected (since energy is defined up to a constant). As a result, the *Clamped Nuclei Hamiltonian* $\hat{H}^{cn}$ turns out to be

$$\hat{H}^{cn} = \sum_{i=1}^{N} \frac{\hat{p}_i^2}{2m} + \frac{e^2}{8\pi\varepsilon_0} \sum_{\substack{i,j=1 \\ i \neq j}}^{N} \frac{1}{|\hat{r}_i - \hat{r}_j|} - \frac{e^2}{4\pi\varepsilon_0} \sum_{g=1}^{A} \sum_{i=1}^{N} \frac{Z_g}{|\hat{r}_i - R_g|}. \tag{6}$$

The Schrödinger equation for this new Hamiltonian becomes

$$\hat{H}^{cn}(\hat{r}_i; R_g) \Psi_n^{cn}(r_i; R_g) = E_n^{cn}(R_g) \Psi_n^{cn}(r_i; R_g), \tag{7}$$

where the semicolon indicates that the corresponding function or operator is only a function of the first argument, while the second is a fixed parameter.

For each energy level $n$, a potential function $V_n(R_g)$ can be defined in terms of the energies $E_n^{cn}(R_g)$ computed for all possible nuclear configurations, that is, for all sets of values $R_g$. In $V_n(R_g)$, the coordinates $R_g$ are no longer classical parameters but classical variables, so that the potential function defines a so-called *Potential Energy Surface* (PES) due to the electrons of the molecule. The quantum behavior of the nuclei is then reintroduced by reestablishing the nuclear kinetic energy $\hat{T}_N(\hat{P}_g)$, and by conceiving them as moving in a quantum potential represented by the PES.

Other approximations are involved in the BOA. Born's *Separability Ansatz* assumes that the molecular wavefunction can be approximated by a sum of products between functions $\theta_{n\alpha}(R_g)$ of nuclei positions and functions $\psi_\alpha(r_i; R_g)$ of electron positions with nuclei positions as parameters (Born 1951). The *Adiabatic Approximation* assumes that the rate of change of the function $\psi_\alpha(r_i; R_g)$ with respect to the classical nuclear positions $R_g$ is very low. These points have been deliberately omitted in the above presentation of the BOA because our papers only draw attention to the classical assumption embedded in the procedure of substituting quantum operators with classical parameters—an assumption that lies at the core of the Clamped Nuclei Approximation. The HLT article devotes almost an entire page to justifying that Born's ansatz "is *not* a sum of nuclear $\{\chi_a(x_1)\}$ and electron $\{\zeta_a(x_2)\}$ (tensor) product states" because "$\psi_a(x_1, x_2)$ [our $\psi_n(r_i; R_g)$] is a wavefunction for *both* parts" (2024: 15; emphasis in the original). Moreover, the article spends significant space showing how the Separability Ansatz and the Adiabatic Approximation can be obtained from the assumption "Heavy" (2024: 15-17). It is unclear what role these efforts play in the general argumentation against our thesis regarding the theoretical tensions embodied in the BOA, as we never mentioned either Born's Separability Ansatz or the Adiabatic Approximation. Since our arguments only refer to the Clamped Nuclei Approximation, in the remainder of the present work we will focus on this approximation.

*3.2.- About the disparity of masses*

The argument of the disparity of masses, which appears in the original version of the BOA, is often presented as a justification for the Clamped Nuclei Approximation. In the case of the HLT article it can be read: "'clamped' is understood purely formally […]. It is *as if* the light, electronic subsystem sees the heavy, nuclear subsystem at a fixed value of $x_1$" (2024: 13, emphasis in the original). This argument is mentioned in the HLT article in the context of the modern BOA (not only in reference to the original 1927 version, called "Perturbative Expansion Born-Oppenheimer approximation" in the article): "The basic idea is to use to the high ratio between the electron and nuclear masses to produce trial solutions to a molecular Schrödinger equation" (2024: 5). Or, more explicitly: "The intuitive idea is that the total kinetic energy of the nuclei is small compared to the potential energy of the molecule and the total kinetic energy of the electrons. (This is because a nucleon is far heavier than an electron, and so typically moves far more slowly, and kinetic energy is $mv^2$)." (2024: 6).

If this were the case, the Clamped Nuclei Approximation would be a standard approximation that could be replaced by allowing the nuclear masses to tend to infinity. Under this assumption, the "so-called 'clamped' Hamiltonian" becomes "the sum of the potential energy of the molecule plus kinetic energy of the electrons only" (HLT 2024: 6). The result of letting the nuclear masses $M_g$ increase without limit in the Coulomb Hamiltonian of eq. (3) is the Electronic Hamiltonian $\hat{H}_0$ of eq. (4), which represents the sum of the molecule's potential energy plus the kinetic energy of the electrons only, as the HLT article claims. However, when compared with eq. (6), it becomes apparent that $\hat{H}_0$ *is not the Clamped Nuclei Hamiltonian*, since in eq. (4) the nuclear positions are still represented by quantum operators; thus, the potential due to the interactions between the nuclei is not a constant scalar that can be neglected. In other words, *the Clamped Nuclei Hamiltonian is not the sum of the potential energy of the molecule plus kinetic energy of the electrons*, as claimed in the HLT article.

Someone might wonder why not dispense with the Clamped Nuclei Hamiltonian and work with the Hamiltonian $\hat{H}_0$ obtained by allowing the nuclear mass to approach infinity, so that the discussions about what clamping the nuclei means would be dissolved from the outset. However, this strategy faces an insurmountable obstacle: as pointed out by Sutcliffe and Woolley (2012a) and by Jecko (2014), $\hat{H}_0$ has a purely continuous spectrum $[E_0, +\infty)$ and, as a consequence, has no normalizable eigenvectors (we will come back to this point below). This would mean that molecules would not have bound states: they would not be stable systems. Moreover, "the decomposition of the molecular Hamiltonian into a nuclear kinetic energy operator contribution, proportional to $\kappa^4$, and a remainder does not yield molecular potential energy surfaces." (Sutcliffe and Woolley 2012a: 6). In other words, clamping the nuclei cannot be easily dispensed with on the basis of the high disparity of masses between nuclei and electrons.

Summing up, the Clamped Nuclei Approximation is not an innocent strategy resulting from making the limit $m/M_g \to 0$. On the contrary, it requires the replacement of quantum operators by classical variables and, as a consequence, differs significantly from any limiting procedure. Since the Clamped Nuclei Approximation is essential to the BOA, it is worthwhile to examine it more closely.

*3.3.- The Clamped Nuclei Approximation*

The HLT article never mentions the Clamped Nuclei Approximation, as commonly called in the literature. Instead, it introduces "*Clamped*: molecular nuclei have fixed definite positions and zero kinetic energy." (2024: 7; emphasis in the original), as the assumption supposedly adopted in our papers. If the term "*Clamped*" is interpreted literally, it implies that nuclei are conceived as classical particles and, therefore, they are treated as such throughout the entire development of the BOA, including its final result. On the basis of this interpretation, the HLT article insists that "[a]ll the states relevant to BO are vectors in a Hilbert space in which the Heisenberg uncertainty relations are automatically respected." (2024: 19). Furthermore, it states that "[a] violation of the relation would require a 'wavefunction' sharply peaked in *both* the position and momentum basis simultaneously, […]. They (sic) would only be violated if the model represented electron or nuclei states as other than vectors in Hilbert space: say, as a delta functions in both position and momentum space." (2024: 19-20).

The points raised in the previous paragraph show that the HLT article builds a "straw man" to argue against it, on the basis of an inaccurate reading of our works. In fact, we never adopted an assumption like "*Clamped*". On the contrary, we introduced the Clamped Nuclei Approximation as *the first step of the BOA*, as is standard in the literature. On the basis of the presentation of Subsection 3.1, it is clear that *the quantum behavior of the nuclei is reintroduced at the last step of the BOA*, when their kinetic energy is reestablished by conceiving them as moving in a quantum potential represented by the PES. Therefore, at the end of the BOA procedure, all the resulting wavefunctions are legitimate quantum functions thereby not violating the Heisenberg Principle. Our concern is the introduction of a first step that treats nuclei as classical, a strategy that gives very good results regarding the calculation of energy levels, but is far from neutral when other molecular properties are considered (this point will be elaborated in Section 4, when analyzing the symmetry problem and the practice of quantum chemistry). Therefore, we will not discuss here an assumption like "*Clamped*," which we never adopted (and no one has done it as far as we know).

Although the HLT article does not mention the Clamped Nuclei Approximation, in several places it does refer to the "clamped Hamiltonian". However, it never explicitly and formally explains how such a Hamiltonian is obtained: its characterization as "the sum of the potential energy of the molecule plus kinetic energy of the electrons only" (2024: 6) is not correct, as explained in the previous subsection. Nevertheless, it seems clear that the term 'clamped Hamiltonian' is intended to refer to the Hamiltonian $\hat{H}^{cn}$ of eq. (6), and that the family of the infinite clamped Hamiltonians for all the possible nuclear configurations should be what defines the Potential Energy Surface. Despite this, the HLT article insists that the "parameterized family of 'clamped' Hamiltonians are used as a tool to construct the effective molecular wavefunction" (2024: 8); "the use of the clamped Hamiltonian (or rather, infinity of clamped Hamiltonians) is purely formal, for constructing a useful expansion, and should not be given physical significance." (2024: 19). Unfortunately, there is not a clear argument to support these claims, other than the assertion that in the *final result of the BOA* the nuclei reacquire their quantum character, something that nobody would deny (unless one subscribes to the "*Clamped*" assumption). What is even more surprising, in the light of the explanation of Subsection 3.1, is the claim that "*At no stage in the method does BO involving representing the nuclei to be anything but quantum.*" (HLT 2024: 19; emphasis in the original).

In the case of the original 1927 version of the BOA, the HLT article correctly stresses that "the representational content of a perturbative model should not be conflated with its zeroth order terms" (2024: 11): the zeroth order of the expansion, in which the nuclei are clamped, is a "fictitious system" (2024: 12). However, in the current standard version, there is no perturbative expansion: the Clamped Nuclei Approximation *turns quantum operators into classical parameters*, and this has nothing to do with a limiting procedure. In John Norton's terms, strictly speaking it is not an approximation but an idealization (Norton 2012), since the target system represented by the Coulomb Hamiltonian $\hat{H}$ of eq. (3) is replaced by another system, in this case represented by the Clamped Nuclei Hamiltonian $\hat{H}^{cn}$ of eq. (6). The HLT article asserts that the BOA "is justified in quantum chemical practice […] in terms of stability under de-idealization" (2024: 6). However, when the article faces de-idealization, it only considers the Separability Approximation and the Adiabatic Approximations, but not the Clamped Nuclei Approximation. Thus, the question remains: How to de-idealize a Clamped Nuclei Hamiltonian (or a family of them) without turning classical variables back into quantum variables "by hand"?

Let us emphasize again that the Clamped Nuclei Approximation is not a strategy similar to those of the traditional semi-classical approximations, which commonly rely on expansions in terms of the ratio of some magnitudes or on the limit $\hbar \to 0$ (strictly speaking $\hbar/S \to 0$, where $S$ is the action of the system).[5] When a limit is involved in the approximation, the role of singular limits as an obstacle to reduction has been widely discussed (see, e.g., Batterman 2001, Rueger 2000a, 2000b, Steeger and Feintzeig 2021). But in the Clamped Nuclei Approximation there is no limit involved, so those discussions cannot simply be extrapolated to this case.

It is interesting to note that the HLT article does not mention the difference between factual and counterfactual approximations (Bruer 1982, Rohrlich 1989), explicitly considered in our papers (Martínez González, Fortin, and Lombardi 2019, Fortin and Lombardi 2021). A counterfactual approximation contradicts a postulate of the theory, but it can be legitimately used if it can be replaced by a factual one. For example, in special relativity the counterfactual limit $c \to \infty$ is legitimate because it can be replaced by the factual limit $v/c \to 0$, that is, an approximation for velocities $v$ much lower than the speed of light $c$. Based on this distinction, the Clamped Nuclei Approximation is a counterfactual "approximation" (or better, idealization), as it assigns definite positions and momenta to the nuclei of the molecule. The problem is that it is unclear which factual approximation could replace it. As explained in the previous subsection, neglecting the operator for the kinetic energy of the nuclei is not sufficient to arrive at the Clamped Nuclei Hamiltonian.

*3.4.- About the spectrum of the Hamiltonians involved in the BOA*

In the HLT article it can be read: "Sutcliffe and Woolley (2012) argue that assumptions regarding the *discrete spectra of electronic Hamiltonians* used in BO are unjustified." (2024: 4; emphasis added). This is not an isolated claim, but it is repeated several times: "the problem that they [Sutcliffe and Woolley] raise is one of formal rigour, namely that BO, involves […] assumptions of normalizable eigenfunctions for operators *without purely discrete spectra*." (HLT: 9; emphasis added); "Recall that Sutcliffe and Woolley (2012) claim that *the clamped Hamiltonian in general has a continuous part to its spectrum*." (2024: 20; emphasis added).

---

[5] By contrast to what is said in the HLT article (2024: 25), a quantum description in which the electromagnetic fields are not quantized is not semi-classical for this reason, because this fact is part of non-relativistic quantum mechanics (it *is* necessary to move on to quantum field theory to have quantized fields).

These quotes show that the HLT article completely misrepresents Sutcliffe and Woolley's work. In fact, the problem pointed out by Sutcliffe and Woolley is not that the Electronic Hamiltonian (eq. (4)) or the Clamped Nuclei Hamiltonian (eq. (6)) do not have a purely discrete spectrum or that they have a spectrum with a continuous part. The problem they raise is that *the Electronic Hamiltonian has a purely continuous spectrum* $[E_0, +\infty)$ and, as a consequence, has no normalizable eigenvectors; as already mentioned, this means that molecules would not have bound states. In the authors' own words: "the exact electronic Hamiltonian […] necessarily has a *purely continuous spectrum of energy levels*; there are *no* potential energy surfaces" (Sutcliffe & Woolley 2012a; 2; emphasis in the original). It is on the basis of this wrong reading of Sutcliffe and Woolley's work that the HLT article constructs the idea of a debate about rigor in the formulation of the BOA.

*3.5.- The supposed debate: Sutcliffe and Woolley "versus" Jecko*

In the HLT article it can be read: "they [Sutcliffe and Woolley] suggest that removing these unjustified mathematical idealizations [about the spectrum of the electronic Hamiltonian] requires making use of resources of classical physics together with empirical data introduced 'by hand'. We consider the *response* to Sutcliffe and Woolley (2012) in Jecko (2014)" (2024: 4-5; emphasis added). This quote suggests a debate between the aforementioned authors, an idea reinforced by repeating that Jecko's work is "both an overview of the mathematical literature on BO, and a *response* to various worries raised by Sutcliffe and Woolley (2012)." (2024: 10; emphasis added). However, a close reading of the two papers reveals that such a debate does not exist because they pursue completely different purposes: pointing out the limitation of Born and Oppenheimer's original approach and obtaining the PES in the case of Sutcliffe and Woolley; approximately obtaining the low-lying bound state energy spectrum (if any) of the Coulomb Hamiltonian in the case of Jecko.

In the Introduction of their paper, Sutcliffe and Woolley clearly say: "We aim to show that the potential energy surface does not arise naturally from the solution of the Schrödinger equation for the molecular Coulomb Hamiltonian; rather its appearance requires the additional assumption that the nuclei can at first be treated as classical distinguishable particles and only later (after the potential energy surface has materialized) as quantum particles." (Sutcliffe and Woolley 2012a: 1). According to the authors, the problem of the original BOA (even in its Born-Huang version) is that the Electronic Hamiltonian has purely continuous spectrum (no bound states, no PES). To achieve their aim, they develop their argumentation in the following steps:

- They introduce a change of variables in terms of center-of-mass coordinates and relative coordinates, obtaining $H = T_{CM} + T_{Nu} + T_{el} + V$.

- They define the *translationally invariant Hamiltonian* $H' = T_{Nu} + T_{el} + V$ (resulting from dropping the kinetic energy of the center of mass in $H$).

- They define the *translationally invariant electronic Hamiltonian* $H^{elec} = T_{el} + V$ (resulting from separating off the kinetic energy of the nuclei from $H'$), and they point out that it has a purely continuous spectrum.

- Finally, they obtain the *clamped nuclei-like Hamiltonian* $H(\mathbf{b}, \mathbf{t}^c)_0$, which is $H^{elec}$ evaluated at the nuclear positions represented by $\mathbf{b}$. This drives us essentially back to the original Clamped Nuclei Hamiltonian (apart from the Hughes-Eckart term which is not relevant).

- Since $H(\mathbf{b}, \mathbf{t}^c)_0$ does not have a purely continuous spectrum (in general it has both discrete and continuous parts of its spectrum), the PES can be obtained by solving the Schrödinger equation with $H(\mathbf{b}, \mathbf{t}^c)_0$ for all the possible nuclear configurations $\mathbf{b}$.

Although according to the HLT article, "Sutcliffe and Woolley's reasoning perhaps suggests a version of *Clamped*" (2024: 9), it is clear that this is not the case: as explicitly emphasized by the authors, "once the classical energy surface has emerged, the nuclei are treated as quantum particles" (Sutcliffe and Woolley 2012a: 7).

Jecko's article (2014) has a completely different aim: the author is absolutely not interested in the PES, but focuses on the energy levels of the molecules. In particular, he is concerned about the fact that the clamped-nuclei-like Hamiltonian still has a continuous part of its spectrum. So, he appeals to the projection method proposed by Jean-Michel Combes (1975), which is a well-known strategy for approximately obtaining the discrete energy spectrum levels of non-relativistic Hamiltonians in small energy ranges. On this basis, Jecko develops the following steps:

- First, the *clamped nuclei Hamiltonian* $Q(x)$ is defined for each configuration $x$ of fixed positions of the nuclei. $Q(x)$ is the $H(\mathbf{b}, \mathbf{t}^c)_0$ as defined by Sutcliffe and Woolley (2012a).

- Each $Q(x)$ has a discrete part of the spectrum below the continuous part. So, the second step consists in defining, for each nuclear configuration $x$, a projector $\Pi(x)$ that projects any wavefunction $\psi(y)$ onto the subspace of the eigenfunctions corresponding to the discrete eigenvalues of $Q(x)$ lower than a certain reference value $E_+$.

- Then, a projector $\Pi$ is defined as the direct integral of the $\Pi(x)$.

- The final step is to define an *effective "adiabatic" Hamiltonian* as $H_{eff} = \Pi H \Pi$, where $H$ is the Coulomb Hamiltonian: $H_{eff}$ has discrete spectrum because all the spectral subspaces of the operators $Q(x)$ corresponding to energies above $E_+$ were removed. The assumption is that the eigenvalues of $H_{eff}$ are good approximations to the eigenvalues of the Coulomb Hamiltonian $H$.

It is interesting to note the three following points. First, the projection method was not developed by Jecko in order to "respond" an argument by Sutcliffe and Woolley, but is a strategy very well-known in the scientific field since the seventies. Second, the projection method gives adequate results for a certain range of energy (the energy range of a stable molecule), which needs to be presupposed in advance on the basis of empirical knowledge. Third, the Hamiltonian $Q(x)$ is also the result of "clamping" the nuclei, so the projection method also appeals to the Clamped Nuclei Approximation. Therefore, claiming that, in Jecko's paper, "mathematical idealizations can be justified without recourse to classical or empirical assumptions." (HLT 2024: 4; emphasis added) is not correct.

This summary presentation of Sutcliffe and Woolley's and Jecko's articles shows that the "debate" between the authors is an artificial construction based on a "biased" understanding of the proposals. A careful reading of those works shows that they approach the BOA from different and compatible perspectives, resulting from the different interests of the authors.

*3.6.- What are chemists interested in?*

In its opening sentence, the HLT article states that "[q]uantum chemistry is the use of quantum mechanics (and quantum field theory) to model molecules and their dynamics, with the goal of explaining and predicting their chemical properties and reactions." (2024: 2). This characterization clearly indicates that the goals of quantum chemistry are chemical goals.[6] Therefore, the question about what kind of knowledge is relevant for chemists is a legitimate one.

In any of its versions, the BOA is a useful strategy for computing energy levels of molecules, which are then tested by spectroscopic measurements. This knowledge is, of course, valuable in the field of chemistry. But, as Hans Primas points out, "quantum mechanics gives perfect predictions for all spectroscopic experiments. However, chemistry is not spectroscopy." (1980: 105). In Sutcliffe and Woolley's words: "the solutions of the time-independent Schrödinger equation for the molecular Hamiltonian are of limited interest for chemistry, being really only relevant to a quantum mechanical account of the physical properties (mainly spectroscopic) of atoms and diatomic molecules in the gas-phase." (2012a: 7).

---

[6] As currently emphasized, quantum chemistry is an in-between, inter-field discipline (Gavroglu and Simões 2012), which combines classical structure theory, quantum concepts, and mathematical approximation techniques (Woody 2012).

Chemists' main interest is not the energy spectrum of molecules but their *molecular structure*. The structure of molecules is what accounts for significant chemical phenomena, such as dipole momenta, functional groups, chemical bonds, isomerism, and optical activity, among others. As John Ogilvie rhetorically asks: "Which is more important to the chemist, quantum mechanics or the concept of molecular structure?" (1990: 286). The central role played by molecular structure in chemistry is explicitly emphasized by many authors: it is "the central dogma of molecular science." (Woolley 1978: 1074); "[t]he alpha and omega of molecular chemistry is the doctrine that molecules exist as individual objects and that every molecule has a shape, characterized by its molecular frame." (Primas 1994: 216; see also Primas 1981); "molecular structure is so central to chemical explanation that to explain molecular structure is pretty much to explain the whole of chemistry." (Hendry 2010: 183); "without the classical concept of molecular structure there is no chemistry" (Woolley 2022: 417).

Given the central role played by the concept of molecular structure in chemistry, the debate about the reduction of chemistry to physics must focus on this concept. Indeed, our claims about the incompatibility between the Clamped Nuclei Approximation and the Heisenberg principle are enunciated in the context of the discussion about the problem of molecular structure, that is, the question about the possibility of explaining molecular structure exclusively in quantum terms, a subject that is scarcely addressed in the HLT article. This article entirely removes the criticized statement from the argumentative context in which it plays a role in our works. Thus, in the next section we will redirect the discussion about the BOA to the framework to which it belongs.

## 4.- What are we talking about?

The HLT article talks about the BOA as an approximation for computing the energy levels of molecules, while in our works we talk about molecular structure in the context of the problem of reduction. In this context, the first step is to clarify the concept of molecular structure.

Following Hendry (2013, 2016, 2021), we will distinguish between two kinds of molecular structure: geometrical structure and bond structure. The *geometrical structure* of a molecule is defined by the relative spatial relations between its component ions or atoms. The *bond structure* of a molecule is the network of bonds connecting its constituent ions or atoms. In what follows, we will only consider molecular structure in its geometrical meaning.[7]

*4.1.- The problem of symmetry*

As advanced in Subsection 2.4, since the interactions between the components of the molecule are Coulombic, the Coulomb Hamiltonian of eq. (3) exhibits certain symmetries with respect to the nuclei: rotation symmetry, reflection symmetry, and permutation symmetry for identical nuclei. The eigenfunctions of this Hamiltonian (in particular, the ground state wavefunction) mathematically inherit those symmetries. However, in many cases, the asymmetry of real molecules is essential to explain their chemical behavior. The *symmetry problem* arises because those relevant asymmetries cannot be explained in terms of the Coulomb Hamiltonian. This problem is independent of the BOA since it stems from the Coulomb Hamiltonian without any approximation.

Hendry illustrates the symmetry problem with the example of the hydrogen chloride molecule, which has an asymmetrical charge distribution that explains its acidic behavior and its boiling point. However, according to quantum mechanics, the expectation value of the electric dipole moment of the molecule in any arbitrary eigenstate of the Coulomb Hamiltonian is always zero: "the acidic behaviour of the hydrogen chloride molecule is conferred by its asymmetry", but "the asymmetry is not conferred by the molecule's physical basis according to physical laws" (Hendry 2010: 186). Another example is the non-zero electric dipole moment of the ammonia molecule: since no stationary state of the molecule can have a nonzero electric dipole moment, the molecule is typically described as "oscillating" between two asymmetric states in a—non-quantum—process known as "umbrella inversion." Optical isomerism is another well-known example: optical isomers, also known as enantiomers, are chiral molecules, meaning that each member of an enantiomeric pair is non-superimposable on its mirror image. The problem arises because the Coulomb Hamiltonian is symmetric under spatial reflection and, as a consequence, its eigenfunctions cannot distinguish between the two members of the enantiomeric pair (for a formal treatment of the latter two cases, see Fortin, Lombardi, and Martínez González 2018, Fortin and Lombardi 2021).

Some authors consider environment induced decoherence (e.g., Zurek 1981, 1991, 2003) as the phenomenon that accounts for the classical limit of quantum mechanics, thereby providing a bridge between the classical concepts of molecular chemistry and the quantum domain (Trost and Hornberger 2009, Scerri 2011, 2013). However, many authors agree that decoherence, by itself, does not solve the problem of the classical limit (see, e.g., Healey 1995, Bub 1997, Pearle 1997, Joos 2000, Adler 2003, Zeh 2003, Schlosshauer 2004, 2007, Bacciagaluppi 2020). We discussed this issue in detail in the case of enantiomers (Fortin, Lombardi, and Martínez González 2016), but the conclusions can be extended to other cases of symmetry. We agree with Pierre Claverie and Simon Diner when they say that "recovering a classical molecular structure amounts to justifying some "classical limit" for the nuclear motions." (1980: 64), but the problem of the classical limit remains "essentialIy unsolved at the present time." (1980: 79). The classical limit of quantum mechanics must be distinguished from what they call "semi-classical limit" (1980: 57), a collection of semi-classical strategies as those mentioned in the last section of the HLT article. Although these strategies are extremely useful in scientific practice, they do not address the conceptual and foundational problem of the quantum-to-classical transition.

---

[7] The obstacles to explaining bond structure in quantum terms are even stronger than those to explaining geometrical structure (see the detailed work of Goodwin 2022).

In the context of a growing sympathy for an "open systems view" of quantum mechanics (Cuffaro and Hartmann 2024; for a critical view, see Lombardi 2025a), the HLT article proposes to transfer such a view to quantum chemistry. In this context, Vanessa Seifert (2022) conceives the use of non-symmetric Hamiltonians as a way to encode persistent interactions between the molecule and the environment. In general, these claims are highly generic, as the specific interactions capable of breaking the symmetry of the isolated molecule are not precisely defined. For instance, many pharmacological drugs are chiral, and usually only one enantiomer exhibits biological activity. This is because protein molecules are chiral, leading to a different reaction with each enantiomer of a chiral drug molecule. However, this raises the question of explaining the chiral nature of biological molecules, pushing the problem a step back. As Hendry observes: "The particular form of the symmetry-breaking addition must be justified however, and it is quite mysterious how that could work if all one has in the environment are more molecules described by Coulombic Hamiltonians. The Coulomb Schrödinger equation for an *n*-molecule ensemble of hydrogen chloride molecules has precisely the same symmetry properties as a Coulomb Schrödinger equation for a 1-molecule system. If the particular form of the symmetry-breaking addition is not justified, then it is just ad hoc: a *deus ex machina*." (Hendry 2010: 186).

In some of our works we have shown in detail how certain particular asymmetric environments can break the symmetry of the Coulomb Hamiltonian of the molecule. For example, in the case of enantiomers, the definite value of the molecule's "chirality" observable, with eigenvalues *d* (dextro-rotation) and *l* (levo-rotation), can be explained on the basis of the interaction between the molecule and polarized light (Fortin, Lombardi, and Martínez González 2018). Similarly, for the ammonia molecule, the electric dipole moment in a specific direction of space, conceived as a quantum observable, acquires a definite value when the molecule interacts with an asymmetric electric field in that spatial direction (Lombardi 2025b).

If the symmetry problem is not solved, then the molecular structure problem is not solved either. However, this does not imply that solving the symmetry problem automatically solves the molecular structure problem. Indeed, symmetry breaking only accounts for the definite value of the observable associated to that symmetry, but it does not explain yet the geometrical disposition of the nuclei in space necessary to define geometrical molecular structure. There are many possible nuclear configurations that realize, say, a left-handed molecule or a molecule with electric dipole moment in a particular direction (for a detailed argument, see Fortin and Lombardi 2021).

*4.2.- The problem of isomers*

Although the problem of isomers is a particular case of the symmetry problem, it introduces additional challenges, especially in the case of structural isomers. As Woolley claims, "the existence of isomers, and the very idea of molecular structure that rationalizes it, remains a central problem for chemical physics." (1998: 3; see also Woolley 2022: Chapter 12).

In a recent article, Alexander Franklin and Vanessa Seifert (2024) argue that "the ground state wavefunction of the resultant [Coulomb] Hamiltonian corresponds to a superposition of *all* the different isomers" (2024: 38-39; emphasis added), each with its own geometrical configuration: quantum measurements lead "to the selection of just one of these configurations and thus account for the transition from symmetric to asymmetric descriptions." (2024: 45). It is true that, in order to recover the symmetry of the ground state from the superposition of asymmetric wavefunctions, *all mathematically possible isomeric structures*, that is, nuclear configurations, must be included in the superposition *with the same probability*. In fact, it is only by superposing all possible asymmetries with equal weight that a symmetry can be restored. This theoretical fact should be empirically manifested by obtaining—at least, approximately—all the possible nuclear configurations with the same frequency in effective experiments. The problem is that this is not the case: given the number and the types of the elements composing a molecule, in general only a few nuclear configurations (when their rotations and translations are neglected) are experimentally observed, that is, only a few isomers are effectively "real."

Of course, there are many *chemical* reasons for the existence of certain isomers and not of others. However, from the viewpoint of quantum mechanics, the question is: why are certain isomeric structures obtained in the laboratory, and not all those necessary to reconstruct the symmetry of the ground state? This question is at the core of the problem of molecular structure, and it is not answered by solving the quantum measurement problem. Furthermore, the stability of molecules that are not in the ground state of the Coulomb Hamiltonian cannot be easily explained exclusively in quantum terms; "How can we explain that, in the course of time, these [initially localized] states retain their localized character? […] even if we are prepared to concede that the "localized" state, different from the molecular eigenstates, has been created by this interaction, […] the molecular Hamiltonian alone is unable to maintain this "localized" character in the course of time." (Claverie and Diner 1980: 68).

*4.3.- The symmetry problem in the scientific literature*

The symmetry problem, with its particular manifestation in the case of isomers, has been recognized by quantum chemists for many decades. Although many claims can be found in Sutcliffe and Woolley papers, here we will mention other authors. For example, regarding the problem of isomerism, in one of his last papers Löwdin stated: "the Coulombic Hamiltonian […] does not provide much obvious information or guidance, since there is (sic) no specific assignments of the electrons occurring in the systems to the atomic nuclei involved—hence there are no atoms, isomers, conformations, etc." (1989: 2071). In a paper devoted to non-BOA calculations for $H_3^+$, Mauricio Cafiero and Ludwik Adamowicz stress that, if nuclei are considered indistinguishable, it is impossible to determine if the molecule is linear or planar triangular: "The only way to predict from the nonadiabatic calculations the actual structure of the molecules and determine whether they are linear or planar triangular is to consider [that] the indistinguishability of the nuclei plays no role" (2004: 139). Very recently, Lucas Lang, Henrique Cezar, Adamowicz, and Thomas Pedersen stated that "understanding the emergence of a purely classical structure, e.g., the fact that chiral molecules exist in unique enantiomers showing optical activity, is still an open problem that remains to be solved." (2024: 1763). Some authors are even more explicit: "molecular structure is a classical concept, and should be represented by classical observables." (Primas 1981: 335); "Therefore, to seek a quantum-mechanical explanation of molecular structure is logically inconsistent." (Ogilvie 1990: 286).[8]

These quotes exemplify the fact that quantum chemists are well aware of the incompatibility between molecular chemistry and quantum mechanics. The question now is how they manage to work within this framework and what role the BOA plays in the actual practice of quantum chemistry.

---

[8] We have also pointed out that the ontology of quantum chemistry is deeply different than that of standard quantum mechanics due to contextuality, non-separability, and indistinguishability (Martínez González, Fortin, and Lombardi 2019). We have even suggested the possibility that Bohmian Mechanics provides a more adequate ontological picture for quantum chemistry (Fortin, Lombardi, and Martínez González 2017, Fortin and Lombardi 2025).

*4.4.- The actual practice of quantum chemistry*

The HLT article discusses the BOA as if its only scientific application were the calculation of the energy levels of molecules. However, the BOA is extensively used in quantum chemistry with specific chemical objectives. Its central role lies in the construction of the Potential Energy Surface (PES), as explained in Subsection 3.1. The PES is one of the fundamental concepts of quantum chemistry as it provides insights into molecular structure and chemical reactions (see Sutcliffe and Woolley 2013).

Let us recall that the PES represents a potential, which is function of the positions of the nuclei and is conceived as being due to the electrons of the molecule. The points of the PES can be classified according to the first and second derivatives of the energy with respect to the nuclear coordinates. The points where the first derivative is zero are *stationary structures*. A stationary structure is a *minimum* if the second derivative along any of the coordinates is positive; it corresponds to an *equilibrium molecular structure* (EQ) that describes a stable chemical species. A stationary structure is a *saddle point* if the second derivative along only one of the coordinates is negative, and along the rest of the coordinates is positive; it corresponds to a *transition state structure* (TS), which connects the basin corresponding to a reactant species to another basin corresponding to a product species through a *minimum energy path*, which is the lowest energy pathway connecting the reactant to the product. In addition, a minimum energy path leading to fragment species is known as a *dissociation channel* (DC). The network including all the EQ, TS and DC is the *global reaction route map* for a given set of nuclei and electrons (see, e.g. Mezey 1987, Schlegel 2003, Harabuchi et al 2015; for a clear conceptual explanation, see Ohno and Satoh 2023).

Through the exploration of the ground-state PES for a certain group of atoms, the preferred geometrical structures, given by the local and the global minima EQ, can be identified. In addition, chemical reactions can be rationalized through the pathway between two minima. For example, the stationary points of the surface lead to the identification of different isomers and the transition structures for conversion between isomers. Besides ground-state PESs, excited-state PESs can also be defined: they are relevant in photochemistry, where some surfaces can come closer to or into contact with one another at points known as *conical intersections*. In fact, whereas thermal reactions stay on the ground-state PES, photochemical reactions can take place in other PESs in different ways: if the photoreaction completely remains on one surface, it is called adiabatic; if the energy surface changes during the reaction, it is called diabatic (see, e.g., Gauglitz 2003).

Although the study of PESs provides chemically significant information about known molecules, its most relevant application is the search for new chemical species and new kinds of reactions: "the identification of new EQ structures (EQs) can allow the discovery of formerly unknown chemical structures, while the evaluation of new TS structures allows novel reaction paths to be elucidated. Also, finding a new DC is equivalent to discovering a new fragmentation mechanism, which is essentially the reverse of an unknown synthetic path." (Ohno and Satoh 2023: 4).

The construction of a PES is a subtle matter. An analytically derived expression for the energy as a function of the nuclear positions can be obtained only for extremely simple chemical molecules. In the generic case, however, the analytical solution is not available, and the strategy is to calculate only the energy for a finite number of nuclear configurations (points on the PES) and to use some computational method of interpolation. Whereas there is an extensive literature on the methods for obtaining a PES (see, e.g. Friesecke and Theil 2015), the important point to emphasize here is that always some configurations must be picked out as the starting point of the computational process: there is no recipe for selecting these configurations, but they are chosen on the basis of chemical experience and intuition.

This brief summary of the central role played by the concept of PES in quantum chemistry explains why Sutcliffe and Woolley are so interested in such a concept in many of their works, in particular in their 2012a paper, which begins by saying: "The principal aim of much of contemporary quantum chemical calculation is to calculate a potential energy surface from solutions of the Schrödinger equation for the clamped nuclei electronic Hamiltonian" (2012a: 1). The HLT article, which only considers the energetic side of the BOA, overlooks this key role, and for this reason focuses on Jecko's paper, where the PES plays no role: in this more physical approach to BOA, the only aim is to approximately obtain the energy levels of a molecule.

Summing up, in *the actual practice of quantum chemistry*, the use of the BOA, with its Clamped Nuclei Approximation, goes far beyond the calculation of energy levels, as it is the basis for understanding and predicting the geometrical structure of molecules and the ways in which this structure is modified through chemical reactions. In the context of this scientific practice, "[t]he most that it seems possible to say is that chemical structure can be teased out in terms of a chosen ansatz (the imposition of fixed, distinguishable nuclei) if one has a good idea of what one is looking for. It seems unlikely that one would ever guess that it was there in the full Coulomb Hamiltonian for a molecular formula unless one had decided on its presence in advance." (Sutcliffe and Woolley 2015: 53). Those classical elements, whose presence is decided in advance, not only lack quantum justification, but are incompatible with the principles of quantum mechanics, in particular, with the Heisenberg Principle.

## 5.- Closing remarks: behind the scene

It seems quite clear that the HLT paper is framed in a hierarchical and unified vision of science, in which fundamental physics plays a foundational role and, as a consequence, potentially explains all scientific knowledge, in particular, chemical knowledge. For at least 40 years, many philosophers of science have been challenging this hierarchical view. However, we have not stated our dissent in these general terms. On the contrary, we have analyzed in detail the argumentative strategies and the technical matters to which the HLT article appeals to support its position. It is on the basis of such an analysis, which takes into account the actual practice of chemistry—which is not confined to computing energy levels and making spectroscopic experiments—, that we have drawn our anti-reductionist conclusions regarding the relationship between chemistry and quantum mechanics. We hope that these arguments will serve to encourage the remarkable philosophers of physics who authored the HLT article to reconsider their conclusion that "the anti-reductionist arguments in philosophy of chemistry are not only unwarranted and incorrect but misidentify the salient challenge to the status of the BO." (HLT 2024: 24), and to reevaluate the aim of setting "an agenda for work in the philosophy of quantum chemistry that is more solidly grounded in scientific practice." (HLT 2024: 1, Abstract).

**Acknowledgements**: We are very grateful to Guy Woolley for his support regarding technical matters. We also want to thank the community of philosophers of chemistry for the interesting and fruitful discussions developed for many years. This work was partially supported by grants PIP 11220200100483CO from CONICET, and PICT-2020-SERIA-00782 from FONCyT. This article is devoted to the memory of Brian Sutcliffe.